\documentclass[]{aastex701}
\usepackage{placeins}

\usepackage{amsmath,amssymb,amsfonts}
\usepackage{CJKutf8}
\usepackage{mathrsfs}
\usepackage{xcolor}

\shorttitle{Processed chemical signatures in polluted white dwarfs}
\shortauthors{Huang et al.}

\begin{document}
\begin{CJK*}{UTF8}{gbsn} 

\title{A Calibrated Bayesian Search for Potential Chemical Technosignatures in Polluted White Dwarfs}

\correspondingauthor{Tong-Jie Zhang, Zhen-Zhao Tao}
\email{tjzhang@bnu.edu.cn, taozhenzhao@dzu.edu.cn}

\author[orcid=0000-0002-8719-3137,sname='Huang']{Bo-Lun Huang(黄博伦)}
\affiliation{Institute for Frontiers in Astronomy and Astrophysics, Beijing Normal University, Beijing 102206, China}
\affiliation{School of Physics and Astronomy, Beijing Normal University, Beijing 100875, China}
\email{Bolunh@hotmail.com}

\author[orcid=0000-0002-4683-5500,sname='Tao']{Zhen-Zhao Tao(陶振钊)}
\affiliation{College of Computer and Information Engineering, Dezhou University, Dezhou 253023, China}
\email{taozhenzhao@dzu.edu.cn}

\author[orcid=0000-0002-3363-9965,sname='Zhang']{Tong-Jie Zhang(张同杰)}
\affiliation{Institute for Frontiers in Astronomy and Astrophysics, Beijing Normal University, Beijing 102206, China}
\affiliation{School of Physics and Astronomy, Beijing Normal University, Beijing 100875, China}
\email{tjzhang@bnu.edu.cn}

\begin{abstract}
 We present a meteorite-calibrated Bayesian framework for searching archival abundance records for \emph{chemical technosignatures}---operationally, compositional patterns better explained by an idealised ``processed'' template (endmember) than by the empirical distribution of natural rocks. We fit a multi-modal natural-composition reference using 3{,}493 whole-rock meteorite analyses, and for each of 697 star--paper abundance sets---spanning at least 397 distinct objects once Gaia-designated repeats are consolidated---we compare the Bayesian evidence for (i) natural material and (ii) a mixture of natural material with a fixed siderophile-enriched template, parameterised by a Ca-normalized mixing fraction $\alpha$. Strong support for the processed template is uncommon: in the photospheric compilation (\texttt{atm}) 8/697 records have $\mathrm{BF}>10$ (4/697 have $\mathrm{BF}>100$), and in the \textcolor{blue}{diffusion-adjusted steady-state subset} (\texttt{acc\_ss}; 148 records spanning at least 94 objects) 6/148 have $\mathrm{BF}>10$. We report the highest-evidence candidate records and infer the fraction of records detectably favoring the mixture model, with posterior medians $\tilde{\pi}=0.011$ (\texttt{atm}) and $\tilde{\pi}=0.041$ (\texttt{acc\_ss}). We calibrate the analysis with end-to-end injection--recovery experiments matched to each record's coverage and censoring. The calibration shows that discrimination is driven mainly by chemical information, typically requires $\gtrsim 5$ detected elements for decisive support, and---for the siderophile template---is strongest for exact five-element panels that include Fe, Mg, Cr, and Ti together with Ni, Si, or Na. These results constrain the detectable incidence of the tested processed-composition class in current data and set observational requirements for future multi-element surveys and expanded template families.
\end{abstract}

\section{Introduction}
\label{sec:intro}

When a Sun-like star exhausts its nuclear fuel, its planetary system enters a second act. Planets and minor bodies can survive post--main-sequence evolution, and long after the host becomes a white dwarf (WD), gravitational perturbations can scatter remnant rocks onto star-grazing orbits. Some of these bodies are tidally disrupted into debris that can form circumstellar dust and/or gas and ultimately accrete onto the WD \citep{Jura2003TidallyDisruptedAsteroid,Gansicke2006GaseousMetalDisk,Farihi2016DebrisPollutionReview}. In the high gravity of a WD, heavy elements gravitationally settle out of the photosphere on timescales far shorter than WD cooling ages \citep{Zuckerman2003MetalLinesDA,Koester2009AccretionDiffusion}. The presence of metals in an otherwise H/He atmosphere therefore demands recent or ongoing external replenishment \citep{Koester2009AccretionDiffusion,Farihi2016DebrisPollutionReview}. Because the accreted material can be decomposed spectroscopically into multiple elements, externally polluted WDs provide unusually direct access to the bulk composition of extrasolar rocky debris \citep{JuraYoung2014,Klein2011,Zuckerman2007ExtrasolarMinorPlanet}. In this sense, polluted WDs are not merely stellar remnants---their atmospheres function as the Universe's only ``exoplanetary autopsy table,'' enabling multi-element assays of destroyed planetesimals and planetary fragments that are otherwise inaccessible in exoplanet observations \citep{JuraYoung2014,Farihi2016DebrisPollutionReview}.

This forensic capability has already made polluted WDs central to exogeoscience: abundance vectors spanning refractory, siderophile, and (sometimes) volatile elements allow quantitative tests of differentiation, crust/mantle/core sampling, and volatile processing by comparison to Solar System rocks \citep{JuraYoung2014,Farihi2016DebrisPollutionReview}. It also motivates a more speculative question: can the chemistry of accreted debris carry \emph{chemical technosignatures}, i.e., compositional patterns plausibly produced by technological activity such as industrial refining, selective concentration, or deliberate alloying \citep{Tarter2001SETI,TechnosignaturesWorkshop2018}? At the bulk-composition level, metal refining and alloying naturally motivate Fe--Ni--Cr-rich processed components, alkali aluminosilicate/glassy processing motivates Na--Al--Si-rich material, and sulfide smelting or ore concentration motivates Fe--Ni--S-rich material \citep{Astafurov2021Stainless,LeLosq2021AAS,Zhao2022NickelSulfide}. In the broader search for extraterrestrial intelligence, much of the chemical-technosignature literature has focused on ``living'' exoplanets, where technology could imprint detectable industrial by-products in atmospheres \citep[e.g.,][]{Lin2014IndustrialPollution}. In parallel, other technosignature work has explored macroscale engineering---including megastructure concepts tailored to WDs \citep{SemizOgur2015DysonSpheres}. Polluted WDs offer a complementary domain: instead of searching for contemporaneous emissions or structures, one can search the \emph{solid debris} itself for evidence that its chemistry has been artificially processed, potentially long after any originating civilization has ceased \citep{TechnosignaturesWorkshop2018}.

However, natural processes such as core--mantle differentiation, partial melting, impact processing, volatilisation/condensation, and selective partitioning into metal or sulfide phases can generate strongly fractionated element patterns \citep{JuraYoung2014,Farihi2016DebrisPollutionReview}. A ``processed''-looking abundance pattern is therefore not, by itself, evidence for technology. Secondly, the polluted-WD literature is archival and disparate: different studies report different element panels, upper limits are common, and multiple analyses of the same star can differ systematically \citep{Williams2024PEWDD}. Without an explicit model for natural compositional diversity \emph{and} a calibration matched to sparse, censored, and inhomogeneous measurements, apparent anomalies can be produced by incompleteness and non-uniform uncertainties rather than by genuinely non-natural chemistry.

Here we present, to our knowledge, the first systematic search specifically targeting \emph{Chemical Technosignatures in Accreted Debris} by treating each published polluted-WD abundance set as a calibrated model-comparison experiment. The key methodological step is to replace ad hoc analogies with an empirical baseline learned from laboratory geochemistry. We construct a \emph{Meteorite Natural Reference} from thousands of whole-rock meteorite analyses drawn from the Astromaterials Data System holdings (including MetBase) \citep{AstromatSynthesis,Hezel2025MetBaseAstromat}, capturing the multi-modal distribution of natural rocky compositions in the same ratio space used for WDs. We then compare, record by record, a natural-only hypothesis to a natural-plus-template mixture hypothesis representing a stylised processed composition. Crucially, we do not interpret Bayes factors at face value: we quantify what the heterogeneous literature can and cannot discriminate by performing end-to-end \emph{Injection--Recovery Calibration} experiments that match observed element panels, uncertainty structure, and upper-limit censoring \citep{Williams2024PEWDD}. 

The remainder of this paper is organized as follows. In Section \ref{sec:methods}, we describe the compilation of archival white dwarf data and the construction of an empirical natural reference model using laboratory meteorite analyses, followed by the formulation of the Bayesian mixture framework. Section \ref{sec:results} presents the record-level model comparison results and, crucially, calibrates the interpretability of these signals through end-to-end injection-recovery experiments matched to observational heterogeneity. Finally, in Section \ref{sec:discussion}, we discuss the inferred population prevalence of the tested technosignature class and define the observational requirements for future multi-element spectroscopic surveys.

\section{Data and Methods}
\label{sec:methods}

\subsection{Data sources and preprocessing}
\label{subsec:methods:data}

\subsubsection{Polluted white dwarf abundances and record definition (PEWDD)}
\label{subsec:data_pewdd}

We start from the Planetary Enriched White Dwarf Database (PEWDD), a literature compilation of photospheric metal abundances for WDs interpreted as accreting planetary debris \citep{Williams2024PEWDD}. PEWDD aggregates heterogeneous analyses and preserves bibliographic provenance, which is essential here because different studies of the same star can yield systematically different abundance patterns and uncertainties.

We define a \emph{WD record} as a \textbf{single star--paper abundance set}. A given star may contribute multiple records if it has been analysed by multiple studies. Each record provides a sparse set of element constraints, each of which is either (i) a detection with a measured Ca-normalized ratio or (ii) an upper limit with a reported threshold. Here ``quality control'' means requiring the Ca reference measurement to be present and not itself an upper limit, transforming reported abundances to Ca-referenced log mass ratios, retaining literature upper limits as censoring thresholds, and carrying forward the corresponding ratio-level uncertainty / imputation flags. After this filtering and Ca normalisation, the photospheric dataset comprises \textbf{2,223} Ca-referenced element-ratio constraints spanning \textbf{697} WD records. These 697 records correspond to 640 literature star identifiers and at least 397 distinct objects once Gaia designations are used to consolidate repeated aliases; we nevertheless retain record-level inference because different papers for the same star often provide materially different element panels and systematic choices.

\subsubsection{Natural reference set: meteorite whole-rock compositions (Astromat Synthesis / MetBase)}
\label{subsec:data_meteorites}

To define an empirical natural compositional reference, we use laboratory whole-rock meteorite analyses from the Astromaterials Data System Synthesis interface and its integrated holdings (including MetBase) \citep{AstromatSynthesis,Hezel2025MetBaseAstromat}. We restrict to \textbf{meteorite} samples and \textbf{whole-rock} analyses, excluding mineral separates and other non-bulk preparations, to approximate parent-body bulk compositions. After applying the same Ca-referenced ratio representation used for WDs, the meteorite dataset comprises \textbf{3,493} distinct whole-rock analyses and \textbf{23,353} Ca-referenced element-ratio values. To represent multi-modality in natural rocky compositions, we label each meteorite analysis with a coarse compositional group: \textbf{chondrite}, \textbf{achondrite} and \textbf{other} (including iron and stony-iron meteorites and analyses not cleanly mapped into the two categories above). Group labels are derived from subtype metadata in the Synthesis export.

\subsubsection{Diffusion timescales and steady-state correction}
\label{subsec:data_mwdd}

For the diffusion-corrected track, we use element-by-element gravitational settling timescales as a function of WD atmospheric parameters and atmosphere type, adopting diffusion grids from the Montreal White Dwarf Database (MWDD) \citep{Dufour2017MWDD,Koester2009AccretionDiffusion}. For each WD record with available $T_{\rm eff}$, $\log g$, and atmosphere type, we interpolate the MWDD grids to obtain $\tau_Z$ for each element and $\tau_{\rm Ca}$ for the reference element. Writing $r_Z \equiv \log_{10}(m_Z/m_{\rm Ca})$, the constant-accretion asymptotic limit $t \gg \tau_Z$ gives $m_Z \propto \dot{M}_Z\tau_Z$, so \textcolor{blue}{under the standard one-zone steady-accretion/settling treatment \citep[e.g.,][]{Koester2009AccretionDiffusion,JuraYoung2014}} the Ca-normalized accretion-rate ratio becomes
\begin{equation}
r_{Z,{\rm acc\_ss}} \equiv r_{Z,{\rm atm}} + \log_{10}\!\left(\frac{\tau_{\rm Ca}}{\tau_Z}\right)\,.
\label{eq:accss}
\end{equation}
\textcolor{blue}{Equation~(\ref{eq:accss}) should therefore be read as the steady-state conversion that follows only when accretion has been approximately constant for several relevant settling times and the one-zone settling approximation is adequate; it is not intended to describe build-up or declining phases. The \texttt{acc\_ss} values analysed here are the diffusion-corrected steady-state abundance ratios stored in the working PEWDD-derived table, and the same table stores the corresponding $\log\tau_Z$ and $\log\tau_{\rm Ca}$ values used to audit the transformation.} It is closest to the ongoing-accretion limit, is generally more defensible for H-atmosphere WDs with shorter diffusion times, and should be interpreted more cautiously for He-atmosphere systems with long retention times. For compact notation, we denote this diffusion-adjusted abundance representation as \texttt{acc\_ss} throughout the paper. After requiring MWDD-grid coverage (H: $6000\le T_{\rm eff}/{\rm K}\le 20000$; He: $8000\le T_{\rm eff}/{\rm K}\le 20000$; $7.7\le \log g \le 8.3$) and the same quality-control rules as above, the diffusion-corrected dataset contains \textbf{714} Ca-referenced element-ratio constraints spanning \textbf{148} WD records. We analyse \texttt{atm} and \texttt{acc\_ss} in parallel to assess robustness to this specific diffusion-timescale correction.

\subsection{Ratio-space framework and observation model}
\label{subsec:methods:ratio_obs}

We analyse polluted-WD compositions in Ca-referenced log mass-ratio space. For each numerator element $Z$ we define
\begin{equation}
r_Z \equiv \log_{10}\!\left(\frac{m_Z}{m_{\rm Ca}}\right)\,,
\label{eq:ratio_def}
\end{equation}
where $m_Z/m_{\rm Ca}$ denotes the composition ratio by mass. \textcolor{blue}{If PEWDD reports number abundances relative to the dominant atmospheric species, $a_Z\equiv\log_{10}(n_Z/n_{\rm atm})$ with $n_{\rm atm}=n_{\rm H}$ or $n_{\rm He}$, the mass-ratio value used here is}
\begin{equation}
\color{blue}{r_Z = \left(a_Z-a_{\rm Ca}\right) + \log_{10}\!\left(\frac{A_Z}{A_{\rm Ca}}\right),}
\label{eq:pewdd_to_massratio}
\end{equation}
\textcolor{blue}{where $A_Z$ is the atomic mass. The same transformation is applied to upper-limit thresholds.}

A record provides a sparse set of element constraints, each of which is either (i) a detection with measured $r_{Z,{\rm obs}}$ or (ii) an upper limit with threshold $r_{Z,{\rm ul}}$. For detections we use a Gaussian likelihood in dex,
\begin{equation}
p(r_{Z,{\rm obs}}\,|\,r_Z) = \mathcal{N}\!\left(r_{Z,{\rm obs}};\,r_Z,\,\sigma_{Z,{\rm tot}}^2\right).
\label{eq:det_lik}
\end{equation}
For upper limits we treat the observation as left-censoring at the reported threshold,
\begin{equation}
p(r_{Z,{\rm ul}}\,|\,r_Z) = \Phi\!\left(\frac{r_{Z,{\rm ul}}-r_Z}{\sigma_{Z,{\rm tot}}}\right),
\label{eq:ul_lik}
\end{equation}
where $\Phi$ is the standard normal CDF.

The PEWDD-derived ratio table stores a ratio-level statistical uncertainty for each Ca-referenced measurement, $\sigma_Z\equiv$ per-element ratio-uncertainty values. \textcolor{blue}{In the analysis, this stored quantity is named \texttt{sigma\_stat\_dex\_filled}: it is the reported propagated ratio uncertainty when available and the element-wise imputed value after filling missing uncertainty entries when necessary.} Where a paper supplies numerator-element and Ca uncertainties, these have already been propagated in quadrature into the ratio-space quantity; when a usable ratio-level uncertainty is absent, we impute the per-element median detected-abundance error from the PEWDD compilation and retain an imputation flag. To account for cross-study systematics and additional unmodelled noise introduced by forming ratios, we inflate uncertainties by adding a systematic term in quadrature and enforcing a minimum floor:
\begin{equation}
\sigma_{Z,{\rm tot}} = \max\!\left(\sqrt{\sigma_{Z}^2 + \sigma_{\rm sys}^2},\,\sigma_{\rm floor}\right),
\label{eq:sigma_total}
\end{equation}
We adopt $\sigma_{\rm sys}=0.15$\,dex because the median star-element scatter among repeated PEWDD measurements is $0.15$ dex, and $\sigma_{\rm floor}=0.05$\,dex because it sits just below the 5th percentile ($\approx 0.064$ dex) of positive reported ratio errors in the working table. Conditional on the latent ratio vector, we treat measurement noise as independent across elements; correlations in intrinsic composition are instead captured by the natural-reference model.

\subsection{Element tier used for multivariate inference}
\label{subsec:methods:tiers}

We focus on the high-coverage nine-dimensional ratio vector
\begin{equation}
\mathbf{r} = \left(r_Z\right)_{Z\in\{\mathrm{Al,Cr,Fe,Mg,Mn,Na,Ni,Si,Ti}\}}\,,
\end{equation}
with Ca as the common denominator. \textcolor{blue}{We refer to this nine-element multivariate inference tier as E10, and $N_{\rm det}$ denotes the number of detected numerator elements within this E10 set.} Phosphorus and sulfur are substantially sparser in both the meteorite reference and WD compilation. When P and/or S are present in a record, we treat them as optional univariate constraints (Section~\ref{subsec:methods:natural_cov}) that multiply into the record likelihood; when absent, these terms are omitted.

\subsection{Meteorite-trained natural composition model}
\label{subsec:methods:natural}

\subsubsection{Mixture model and missing-data handling}
\label{subsec:methods:natural_mix}

We model the natural distribution of \(\mathbf{r}\) as a group mixture of multivariate Gaussians:
\begin{equation}
p_{\rm nat}(\mathbf{r}) = \sum_{g=1}^{G} w_g\,\mathcal{N}\!\left(\mathbf{r};\,\boldsymbol{\mu}_g,\boldsymbol{\Sigma}_g\right),
\label{eq:nat_mixture}
\end{equation}
where \(g\) indexes broad meteorite groups (chondrite, achondrite, other), and \(w_g\) are weights estimated from meteorite analysis counts.

Meteorite analyses and WD records do not uniformly report all elements. We therefore evaluate equation~(\ref{eq:nat_mixture}) on the observed subvector for each analysis/record by marginalising each Gaussian component to the observed dimensions. This allows likelihood evaluation under arbitrary element panels without discarding incomplete records.

Importantly, this is not an element-by-element draw from unrelated meteorite classes. Each mixture component carries a full covariance matrix, so element ratios remain coupled in the same way they are coupled within the meteorite data. What the model does \emph{not} attempt is subtype attribution to a single named meteorite class or explicit evolutionary scenario; instead, it asks whether a WD record lies anywhere on the empirical natural manifold spanned by broad rocky-material populations. In that sense the framework is complementary to Bayesian polluted-WD studies that compare individual systems or populations against specific meteorite classes or geological scenarios \citep{Swan2023DZ,Buchan2024Population}.

\subsubsection{Covariance estimation and regularisation}
\label{subsec:methods:natural_cov}

Within each meteorite group, covariances are estimated using pairwise-available meteorite analyses and then regularised by shrinkage toward the diagonal to ensure positive-definite covariance matrices. This stabilises inference when some element pairs have limited overlap in the laboratory dataset. We validate the practical value of the multi-component natural reference via stratified cross-validation (Section~\ref{sec:results:natural_ref}). For \(Z\in\{\mathrm{P,S}\}\), we fit robust univariate normal approximations in Ca-referenced ratio space using meteorite analyses with reported values. When P and/or S are present for a WD record, we multiply the corresponding univariate likelihood terms into the record likelihood; when absent, these factors are omitted.

\subsection{Processed-template mixture hypothesis and Bayes factors}
\label{subsec:methods:endmember}

\subsubsection{Template composition definition}
\label{subsec:methods:template_def}

We test for evidence of a fixed processed \emph{template} composition mixed with natural material. The fiducial template is a \emph{siderophile-enriched concentrate}, but it is best viewed as an \emph{a deliberately stylized endmember used to evaluate detectability} rather than as a unique prediction for technology. In practice, the processed templates are constructed directly from meteorite whole-rock quantiles: enriched elements are set to the 99th percentile plus 0.6--1.0 dex, depleted elements to the 5th percentile, and neutral elements to the median. The fiducial siderophile template therefore represents a metal-rich, silicate-poor component with enhanced Fe, Ni, Cr, and Mn; the three supplementary templates are defined analogously to mimic broad classes of processed material---a refractory-silicate concentrate (Al--Si--Ti rich), an alkali-aluminosilicate concentrate (Na--Al--Si rich), and a sulfide-metal concentrate (Fe--Ni--S rich) \citep{Astafurov2021Stainless,LeLosq2021AAS,Zhao2022NickelSulfide}. Table~\ref{tab:endmembers} lists the Ca-normalized log-ratio values used for the fiducial template and for the supplementary calibration tests. A natural-control template (median chondrite) is included as a negative control.

\subsubsection{Mixing in linear ratio space}
\label{subsec:methods:mixing}

Because log-ratios do not mix linearly, we define mixing in linear Ca-normalized ratio space. Let \(R_Z \equiv 10^{r_Z}=m_Z/m_{\rm Ca}\) and define a Ca-normalized mixing fraction \(\alpha\in[0,1]\). For a natural draw \(\mathbf{R}_{\rm nat}\) and template ratios \(\mathbf{R}_{\rm temp}\),
\begin{equation}
\mathbf{R}_{\rm mix}(\alpha) = (1-\alpha)\,\mathbf{R}_{\rm nat} + \alpha\,\mathbf{R}_{\rm temp},
\label{eq:mix_linear}
\end{equation}
with \(\mathbf{r}_{\rm mix}(\alpha)=\log_{10}\mathbf{R}_{\rm mix}(\alpha)\).

\subsubsection{Record-level evidence and posterior over \(\alpha\)}
\label{subsec:methods:evidence}

For each record we compare:
\begin{enumerate}
\item \(H_0\) (natural-only): \(\alpha=0\),
\item \(H_1\) (natural+template): \(\alpha>0\) with prior \(p(\alpha)={\rm Uniform}(0,1)\).
\end{enumerate}
Under \(H_1\) we marginalise over both the natural composition and \(\alpha\) by Monte Carlo integration over \(\mathbf{r}_{\rm nat}\sim p_{\rm nat}\) and numerical integration over \(\alpha\) on a fixed grid:
\begin{equation}
p({\rm data}\,|\,H_1) \approx \sum_k \left[\frac{1}{N}\sum_{n=1}^{N} p\!\left({\rm data}\,\middle|\,\mathbf{r}_{\rm mix}^{(n)}(\alpha_k)\right)\right] p(\alpha_k)\,\Delta\alpha.
\label{eq:h1_evidence}
\end{equation}
We compute the natural-only evidence \(p({\rm data}\,|\,H_0)\) as the \(\alpha=0\) special case. Support for the mixture model is summarised by the log Bayes factor
\begin{equation}
\ln {\rm BF} \equiv \ln \frac{p({\rm data}\,|\,H_1)}{p({\rm data}\,|\,H_0)}.
\label{eq:lnbf}
\end{equation}
Conditioned on \(H_1\), the posterior over \(\alpha\) is proportional to \(p({\rm data}|\alpha,H_1)p(\alpha)\) \citep{KassRaftery1995,Trotta2008}. For records with upper limits, censoring is handled using equation~(\ref{eq:ul_lik}). When multiple upper limits are present simultaneously, we evaluate the multivariate censored probability by conditioning the relevant Gaussian on the detected subset and estimating the remaining truncated probability by Monte Carlo sampling from the conditional Gaussian, ensuring that records with complex censoring patterns remain usable rather than requiring ad hoc filtering.

\subsection{Population prevalence (detectable incidence)}
\label{subsec:methods:prevalence}

To infer the incidence of template-like records in the ensemble, we use a spike-and-slab prevalence model with parameter \(\pi\in[0,1]\) \citep{MitchellBeauchamp1988}:
\begin{equation}
p({\rm data}_i\,|\,\pi) = (1-\pi)\,p_{0,i} + \pi\,p_{1,i},
\label{eq:pi_mix}
\end{equation}
where \(p_{0,i}\equiv p({\rm data}_i|H_0)\) and \(p_{1,i}\equiv p({\rm data}_i|H_1)\) are record-level evidences. With a uniform Beta\((1,1)\) prior on \(\pi\), we compute the posterior on a dense grid and marginalise to obtain a posterior for \(\pi\) and prevalence-aware slab probabilities for each record. This \(\pi\) is a \emph{record-level} detectable-incidence parameter, not necessarily a unique-star occurrence rate. Mapping it to a per-star quantity would require an additional hierarchical layer that compresses repeated analyses of the same WD; in Section~\ref{subsec:discussion:pi} we therefore report simple star-level sensitivity tests alongside the record-level inference.

\subsection{End-to-end calibration by injection--recovery}
\label{subsec:methods:injrec}

Because the WD literature is heterogeneous in element panels, uncertainties, and censoring, Bayes factors must be calibrated under realistic observational patterns. We therefore perform end-to-end injection--recovery experiments that mirror the complete inference analysis. Synthetic records are generated by:
\begin{enumerate}
\item drawing \(\mathbf{r}_{\rm nat}\sim p_{\rm nat}\),
\item injecting a known template fraction \(\alpha_{\rm true}\) via equation~(\ref{eq:mix_linear}),
\item bootstrapping an \emph{empirical observational pattern} by sampling a real WD record uniformly from the relevant track and inheriting exactly that record's element panel, detection-versus-upper-limit flags, and per-element \texttt{sigma\_stat\_dex\_filled} values,
\item generating detections using equations~(\ref{eq:det_lik})--(\ref{eq:sigma_total}), and generating upper limits as one-sided thresholds
\begin{equation}
r_{Z,{\rm ul}} = r_{Z,{\rm true}} + \kappa\,\sigma_{Z,{\rm tot}},
\label{eq:ul_sim}
\end{equation}
with \(\kappa=2\). This is not an exposure-time or line-transfer model; it is an archival-style censoring surrogate that places the quoted limit about two total-sigma above the truth, i.e. a conservative one-sided ``2\(\sigma\)-like'' upper limit while preserving the observed record-to-record dispersion in element coverage and uncertainties,
\item analysing each injected record identically to real records to obtain calibrated false-positive rates, detection power, mixing-fraction recovery, and prevalence recovery.
\end{enumerate}
For a Bayes-factor threshold \(\eta\), we define the detection power as
\begin{equation}
{\rm Power}(\alpha_{\rm true};\eta) \equiv P\!\left(\ln {\rm BF} > \ln \eta \,\middle|\, \alpha_{\rm true}\right),
\label{eq:power_def}
\end{equation}
with the false-positive rate given by the \(\alpha_{\rm true}=0\) special case. We report power curves both overall and stratified by the number of detected elements \(N_{\rm det}\), which quantifies the effective chemical information content of a record.

\section{Results}
\label{sec:results}

\subsection{Meteorite whole-rock compositions define a structured natural reference}
\label{sec:results:natural_ref}

\begin{figure*}[t]
\centering
\includegraphics[width=0.82\textwidth]{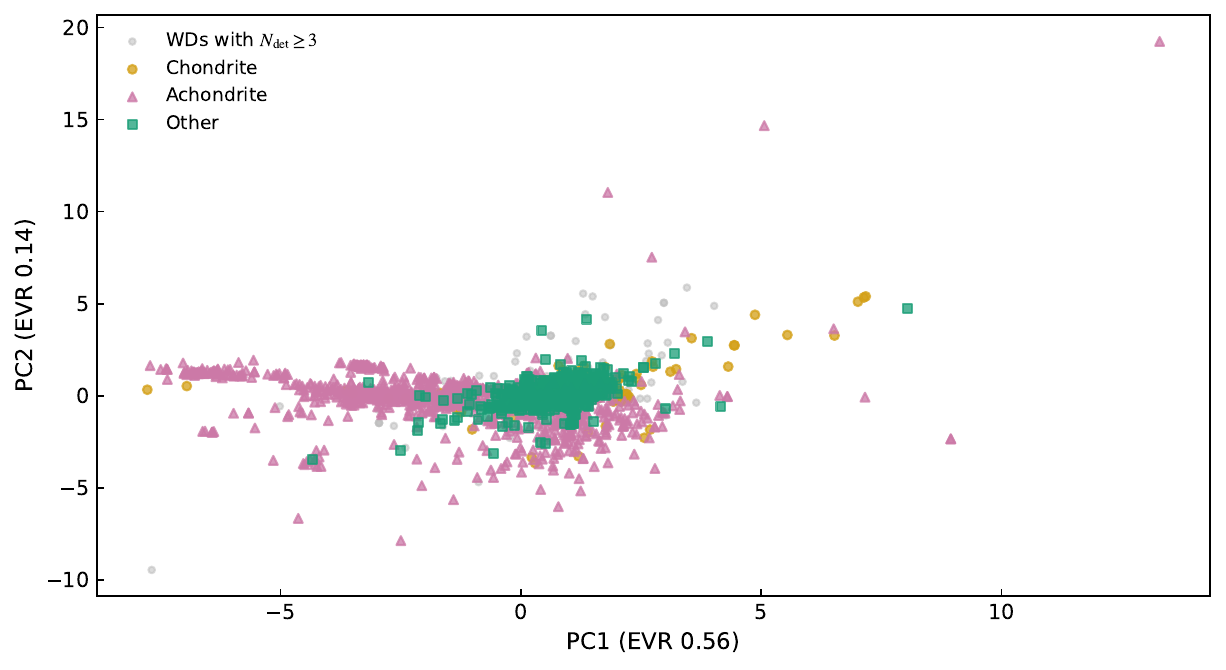}
\caption{\textbf{Natural composition structure in ratio space (meteorites) with white-dwarf overlay.}
Principal component analysis (PCA) projection of Ca-referenced log-ratio vectors derived from whole-rock meteorite analyses, coloured by broad meteorite class (chondrite, achondrite, other). Grey points show polluted WDs with \(N_{\rm det}\ge 3\) projected into the same standardised ratio space. Missing dimensions are median-imputed \emph{for visualisation only}; the likelihood analysis itself marginalises missingness directly. Axes show the first two principal components with explained variance ratios (EVR). \textcolor{blue}{The non-unimodal meteorite support and the fact that most WD projections lie within this broad natural-composition envelope.}}
\label{fig:natural_pca}
\end{figure*}

Whole-rock meteorite compositions occupy a structured, multi-modal region in Ca-referenced ratio space (Fig.~\ref{fig:natural_pca}), motivating a natural reference that can represent multiple composition modes rather than a single unimodal population. \textcolor{blue}{The main feature to focus on in Fig.~\ref{fig:natural_pca} is the broad, structured overlap between the WD projections and the meteorite cloud, rather than the location of any individual point.} For visualisation only, Fig.~\ref{fig:natural_pca} median-imputes missing dimensions and projects the meteorite and WD tables into the first two principal components of the standardised meteorite matrix; the inference itself never uses this imputation or the PCA projection. Up to an arbitrary sign, the first component is dominated by the joint Fe--Mg--Si--Ni--Ti covariance, while the second is driven mainly by an Al+Cr versus Na contrast. The figure should therefore be read as a qualitative map of the learned natural manifold, not as the low-dimensional model used in the likelihood.

\textcolor{blue}{The PCA element groupings are descriptive rather than mechanistic. They arise from the covariance matrix of Ca-normalized whole-rock ratios: the first component tracks a broad metal/silicate abundance gradient relative to Ca, whereas the second captures a residual contrast between refractory/lithophile components and more volatile alkali behaviour.}

\begin{figure}[t]
\centering
\includegraphics[width=0.72\linewidth]{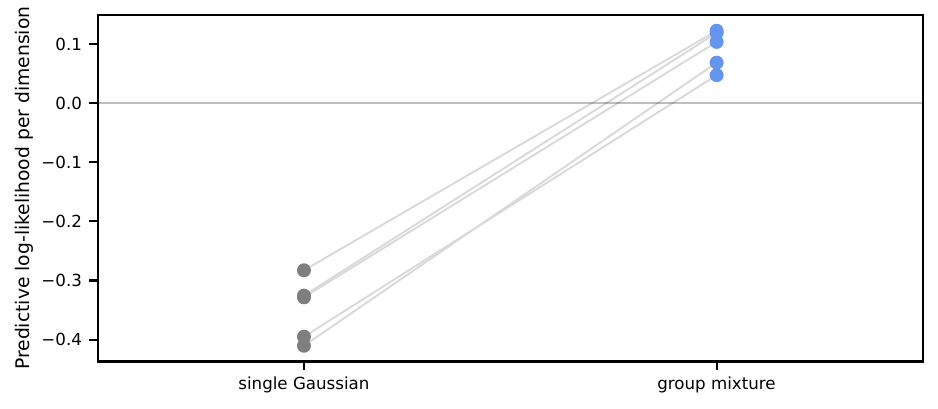}
\caption{\textbf{Cross-validation support for the group-mixture natural reference.}
Predictive log-likelihood per observed dimension evaluated via 5-fold stratified cross-validation. The group-mixture model consistently outperforms a single-Gaussian baseline across folds.}
\label{fig:nat_cv}
\end{figure}

We evaluate the practical value of the multi-component structure using 5-fold stratified cross-validation on meteorite analyses (Fig.~\ref{fig:nat_cv}). The group-mixture natural reference achieves a predictive log-likelihood of $0.092 \pm 0.033$ per observed dimension across folds, compared with $-0.348 \pm 0.053$ for a single multivariate Gaussian fit to the full meteorite set. This improvement supports using a multi-component natural reference when assessing whether polluted-WD compositions are consistent with empirically sampled rocky-material diversity.

This predictive-gain test answers a different question than fitting each WD against named meteorite classes one by one: subtype-by-subtype fits are invaluable for provenance studies \citep{Swan2023DZ,Buchan2024Population}, whereas our aim here is first to determine whether a record is compatible with the broad natural manifold at all before arguing about which natural subtype it most resembles.

\FloatBarrier

\subsection{Element coverage and censoring in the archival WD compilation}
\label{sec:results:coverage}

Element coverage varies strongly across the compiled WD literature records, and this heterogeneity sets the effective discriminating power of any multivariate hypothesis test. Element-level coverage, upper-limit fractions, and missing-uncertainty rates are summarised in Table~\ref{tab:data_coverage}.

In the photospheric track (697 records), the median number of detected E10 numerator elements is \(N_{\rm det}=2\); 382/697 (\(55\%\)) records have exactly two detected elements, and 87/697 (\(12.5\%\)) have \(N_{\rm det}\ge 5\). The \texttt{atm} sample is dominated by He atmospheres (589 He, 108 H) and is relatively cool overall (median \(T_{\rm eff}=6515\) K). The diffusion-corrected subset (148 records) is, by construction, information-richer: the median is \(N_{\rm det}=3\), 45/148 (\(30\%\)) have \(N_{\rm det}\ge 5\), and the stellar sample is warmer and less He-dominated (86 He, 62 H; median \(T_{\rm eff}=11890\) K). The reduction from 697 to 148 records is almost entirely a grid-coverage effect rather than a selection imposed during model comparison: among the photospheric records omitted from \texttt{acc\_ss}, 443 lie outside the adopted He \(T_{\rm eff}\) range, 39 outside the H range, 42 outside the MWDD \(\log g\) range, and 25 lack sufficient atmospheric metadata. Because the records that survive this cut are warmer and generally better studied, the \texttt{acc\_ss} subset is also richer in detected elements and more frequently shows infrared-excess signatures (62/148 versus 94/697 in \texttt{atm}). These empirical detection-pattern and stellar-parameter statistics provide essential context for interpreting record-level evidence, candidate lists, and calibration results.

\begin{deluxetable}{lrrrrrr}
\tablecaption{\textbf{Data summary and element coverage in ratio space.}\label{tab:data_coverage}}
\tabletypesize{\scriptsize}
\tablewidth{0pt}
\tablehead{
\colhead{Element $Z$} &
\colhead{$N_{\rm atm}$} & \colhead{UL$_{\rm atm}$ (\%)} & \colhead{$\sigma$ missing$_{\rm atm}$ (\%)} &
\colhead{$N_{\rm acc\_ss}$} & \colhead{UL$_{\rm acc\_ss}$ (\%)} &
\colhead{$N_{\rm met}$}
}
\startdata
Mg & 670 & 2.4  & 39.4 & 143 & 4.2  & 3128 \\
Fe & 647 & 4.5  & 41.9 & 125 & 10.4 & 3403 \\
Na & 186 & 19.4 & 73.7 & 36  & 61.1 & 2717 \\
Si & 165 & 26.7 & 28.5 & 104 & 26.0 & 2151 \\
Cr & 146 & 15.8 & 57.5 & 54  & 24.1 & 2850 \\
Ti & 108 & 13.9 & 40.7 & 56  & 14.3 & 2306 \\
Al & 99  & 47.5 & 49.5 & 64  & 46.9 & 3091 \\
Ni & 95  & 36.8 & 48.4 & 57  & 38.6 & 417  \\
Mn & 59  & 33.9 & 35.6 & 39  & 25.6 & 3170 \\
S  & 28  & 53.6 & 53.6 & 22  & 50.0 & 58   \\
P  & 20  & 70.0 & 70.0 & 14  & 64.3 & 62   \\
\enddata
\tablecomments{UL = upper limit. Counts are the number of Ca-referenced log mass-ratio constraints for each numerator element \(Z\) in the photospheric track (\texttt{atm}) and diffusion-adjusted subset (\texttt{acc\_ss}), together with the fraction of constraints that are upper limits and the fraction with missing per-ratio statistical uncertainty (requiring imputation; Section~\ref{subsec:methods:ratio_obs}). Meteorite counts give the number of whole-rock analyses contributing \(Z/{\rm Ca}\) ratios in the natural reference ensemble.}
\end{deluxetable}

\FloatBarrier

\subsection{Natural-reference scores identify a low-likelihood tail in both tracks}
\label{sec:results:naturalness}

\begin{figure*}[t]
\centering
\includegraphics[width=0.85\textwidth]{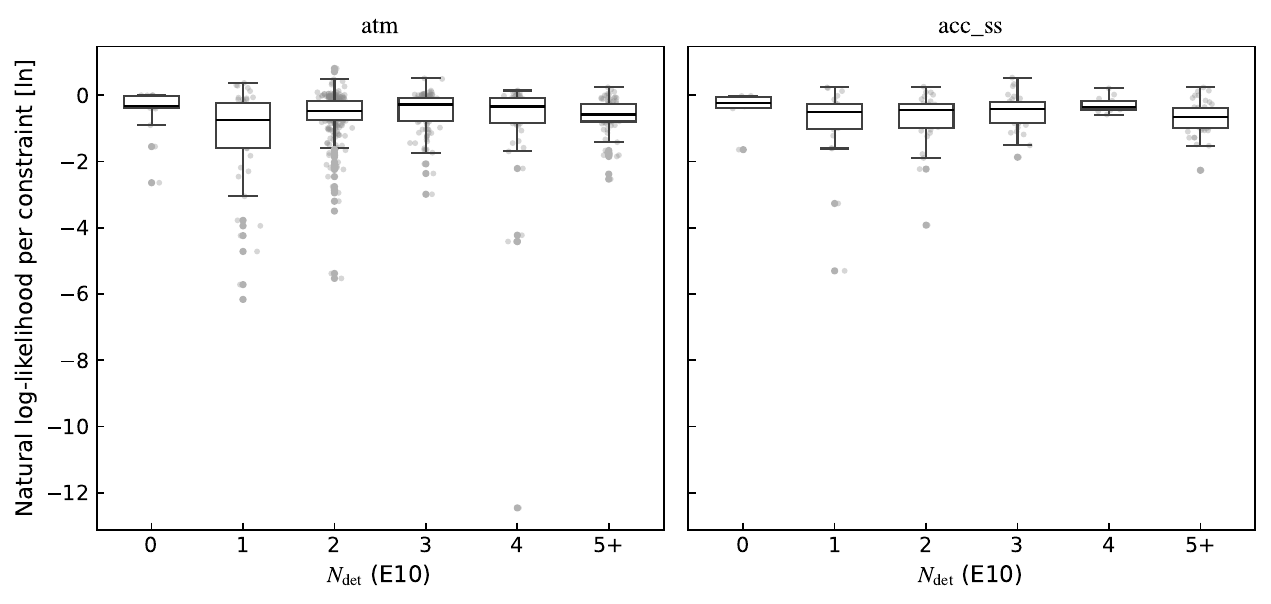}
\caption{\textbf{Natural-reference score versus chemical information content.}
Distribution of per-record natural log-likelihood per constraint, \(\ln p({\rm data}|H_0)/n_{\rm used}\), under the meteorite-trained natural-reference model as a function of the number of detected E10 numerator elements \(N_{\rm det}\) (\(E10\equiv\{\mathrm{Al,Cr,Fe,Mg,Mn,Na,Ni,Si,Ti}\}\)). \textbf{a,} Photospheric track (\texttt{atm}). \textbf{b,} Diffusion-adjusted track (\texttt{acc\_ss}). Boxplots summarise the distribution within each \(N_{\rm det}\) bin; grey points show individual records. Higher values indicate compositions that are more consistent with the natural meteorite reference.}
\label{fig:naturalness}
\end{figure*}

As a baseline diagnostic, we score each WD record under the meteorite-trained natural reference (natural-only hypothesis) and summarise the resulting natural log-likelihood per constraint, \(\ln p({\rm data}|H_0)/n_{\rm used}\), as a function of information content (Fig.~\ref{fig:naturalness}). Two broad trends emerge. First, as \(N_{\rm det}\) increases, the score distribution narrows, consistent with richer element panels more tightly constraining feasible regions of composition space. Second, both tracks exhibit a long low-likelihood tail, indicating that a minority of records are difficult to reconcile with the natural reference given their reported ratios and uncertainties. These scores are therefore best interpreted as a screening statistic: they identify records where explicit model comparison is most informative, but only rich-element panels can make that comparison decisive.

Records with \(N_{\rm det}=0\) are not unconstrained objects. They are upper-limit-only E10 records, with a median of two upper limits in both tracks, so their likelihoods are computed entirely from the censoring terms of equation~(\ref{eq:ul_lik}). The sparse-panel population is dominated by Fe and Mg detections, but the low-likelihood tail is disproportionately associated with Na and Si excursions: among \texttt{atm} records with \(N_{\rm det}\in\{1,2\}\), Na appears \(5.2\times\) more often in the lowest-decile tail than in the parent low-\(N_{\rm det}\) set, and Si appears \(2.1\times\) more often. We therefore view low-\(N_{\rm det}\) outliers as useful follow-up priorities, not as stand-alone technosignature claims.

\begin{deluxetable*}{r l c l c}
\tablecaption{\textbf{Most informative exact five-element panels for the fiducial siderophile template.}\label{tab:best_panels}}
\tabletypesize{\scriptsize}
\tablewidth{0pt}
\tablehead{
\colhead{Rank} &
\colhead{Best \texttt{atm} panel} &
\colhead{Mean power} &
\colhead{Best \texttt{acc\_ss} panel} &
\colhead{Mean power}
}
\startdata
1 & Cr, Fe, Mg, Ni, Ti & 0.852 & Cr, Fe, Mg, Ni, Ti & 0.855 \\
2 & Cr, Fe, Mg, Na, Ti & 0.766 & Cr, Fe, Mg, Si, Ti & 0.671 \\
3 & Cr, Fe, Mg, Si, Ti & 0.728 & Fe, Mg, Na, Ni, Ti & 0.664 \\
4 & Fe, Mg, Na, Ni, Ti & 0.661 & Al, Fe, Mg, Ni, Ti & 0.637 \\
\enddata
\tablecomments{Entries rank exact five-element panels by the mean \(\mathrm{BF}>10\) power over \(\alpha_{\rm true}=0.10\)--0.50, using only panels with at least 50 synthetic realisations in the corresponding track. The repeated appearance of Cr and Ti across the highest-power rows is the main observational guide for the fiducial siderophile-enhancement search.}
\end{deluxetable*}

\FloatBarrier

\subsection{Record-level evidence for a siderophile-enriched processed template is rare}
\label{sec:results:evidence}

\begin{figure*}[t]
\centering
\includegraphics[width=0.9\textwidth]{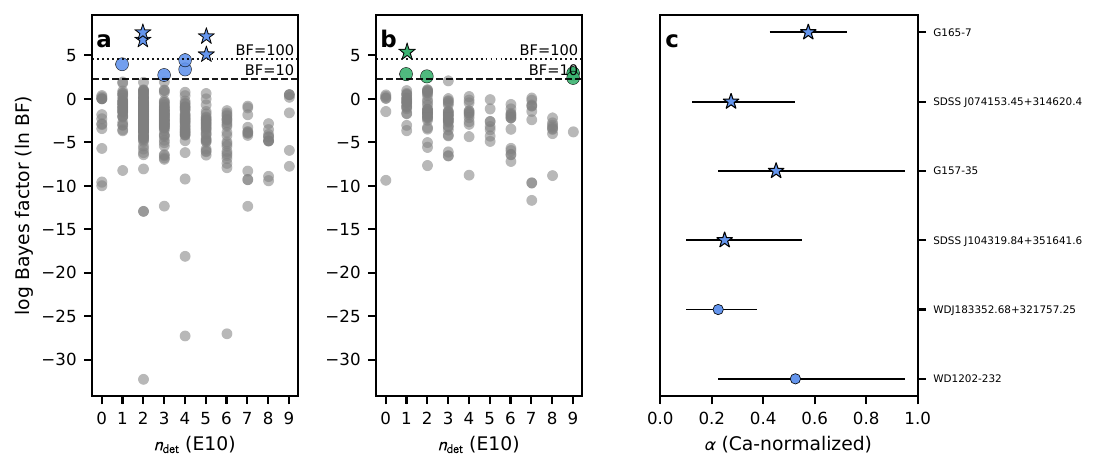}
\caption{\textbf{Evidence for an additional processed template and inferred mixing fractions for highlighted candidates.}
Evidence for a mixture model (natural reference + siderophile-enriched template) relative to a natural-only model shown as \(\ln\mathrm{BF}\) versus \(N_{\rm det}\). \textbf{a,} \texttt{atm}. \textbf{b,} \texttt{acc\_ss}. Horizontal lines mark Bayes-factor thresholds (BF=10 and BF=100). Highlighted markers denote the highest-evidence records in each track. \textbf{c,} For highlighted \texttt{atm} records, posterior median and 95\% credible interval of the Ca-normalized mixing fraction \(\alpha\), conditional on the mixture model.}
\label{fig:endmember}
\end{figure*}

We next apply the processed-template mixture test and quantify support using the record-level log Bayes factor, \(\ln\mathrm{BF}\) (Fig.~\ref{fig:endmember}). In this model, \(\alpha \in [0, 1]\) denotes the Ca-normalized fraction of the fixed template in linear ratio space (\(\alpha = 0\) corresponds to natural-only; Section~\ref{subsec:methods:endmember}). Across the photospheric compilation (697 records), the median \(\ln\mathrm{BF}\) is $-1.964$, so most records prefer the natural-only hypothesis. Only 50/697 (\(7.2\%\)) satisfy \(\ln\mathrm{BF}>0\) (\(\mathrm{BF}>1\)), 8/697 exceed \(\mathrm{BF}>10\), and 4/697 exceed \(\mathrm{BF}>100\); in other words, strong support for the mixture model is uncommon.

The diffusion-corrected subset shows the same qualitative behaviour but with a larger fraction of records weakly favouring the mixture model, consistent with its generally richer element panels: in \texttt{acc\_ss}, 25/148 (\(16.9\%\)) have \(\mathrm{BF}>1\), 6/148 exceed \(\mathrm{BF}>10\), and 1/148 exceeds \(\mathrm{BF}>100\).

\FloatBarrier

\subsection{Highest-evidence candidate records}
\label{sec:results:candidates}

Tables~\ref{tab:candidates_atm} and \ref{tab:candidates_acc} list the highest-evidence candidate records under the fiducial siderophile-template hypothesis. The mixing-fraction column reports the posterior median and 95\% credible interval for \(\alpha\), conditional on the mixture model.

Interpreting candidates requires calibration of detectability under realistic literature patterns. In particular, injection--recovery experiments (Section~\ref{sec:results:calibration}) show that records with sparse detected element panels can yield large \(\ln\mathrm{BF}\) values while still being information-limited: the inferred \(\alpha\) can remain broad and the decision can be sensitive to which elements happen to be reported. Accordingly, sparse-panel candidates are best interpreted as high-priority follow-up targets rather than as robust classifications.

\begin{deluxetable*}{lrcrrlrl}
\tablecaption{\textbf{Candidate WD records in the photospheric (\texttt{atm}) track under the siderophile-template mixture hypothesis.}\label{tab:candidates_atm}}
\tabletypesize{\scriptsize}
\tablewidth{0pt}
\tablehead{
\colhead{Star} & \colhead{$T_{\rm eff}$ (K)} & \colhead{\textcolor{blue}{Atm.}} & \colhead{$N_{\rm det}$} & \colhead{$n_{\rm ul}$} &
\colhead{Detected E10 panel} & \colhead{$\ln \mathrm{BF}$} & \colhead{$\alpha$ (median [95\% CI])}
}
\startdata
SDSS J074153.45$+$314620.4 & 5592 & \textcolor{blue}{He} & 5 & 0 & Cr, Fe, Mg, Ni, Ti & 7.25 & 0.282 [0.120, 0.535] \\
SDSS J104319.84$+$351641.6 & 6720 & \textcolor{blue}{He} & 5 & 0 & Cr, Fe, Mg, Ni, Ti & 5.19 & 0.244 [0.092, 0.544] \\
WDJ183352.68$+$321757.25   & 7650 & \textcolor{blue}{He} & 4 & 0 & Fe, Mg, Ni, Ti     & 4.48 & 0.216 [0.110, 0.372] \\
G165$-$7                   & 7500 & \textcolor{blue}{He} & 2 & 0 & Fe, Mg              & 7.66 & 0.576 [0.412, 0.716] \\
G157$-$35                  & 6000 & \textcolor{blue}{He} & 2 & 0 & Fe, Mg              & 6.84 & 0.458 [0.214, 0.950] \\
WD 1202$-$232              & 8619 & \textcolor{blue}{H}  & 1 & 0 & Fe                  & 4.03 & 0.532 [0.216, 0.936] \\
\enddata
\tablecomments{The first three rows represent the highest-BF candidates with $N_{\rm det}\ge3$, whereas the bottom three represent the highest-BF candidates with $N_{\rm det}<3$. \textcolor{blue}{Atm. gives the dominant atmospheric species (H or He) adopted for the record.} The detected-panel column lists the detected E10 elements $\{\mathrm{Al,Cr,Fe,Mg,Mn,Na,Ni,Si,Ti}\}$ only. Mixing-fraction intervals are conditional on the mixture model. Calibration in Section~\ref{sec:results:calibration} shows that robust discrimination typically requires richer detected element panels; sparse-panel candidates are prioritised follow-up targets.}
\end{deluxetable*}

\begin{deluxetable*}{lrcrrlrl}
\tablecaption{\textbf{Candidate WD records in the diffusion-corrected (\texttt{acc\_ss}) track under the siderophile-template mixture hypothesis.}\label{tab:candidates_acc}}
\tabletypesize{\scriptsize}
\tablewidth{0pt}
\tablehead{
\colhead{Star} & \colhead{$T_{\rm eff}$ (K)} & \colhead{\textcolor{blue}{Atm.}} & \colhead{$N_{\rm det}$} & \colhead{$n_{\rm ul}$} &
\colhead{Detected E10 panel} & \colhead{$\ln \mathrm{BF}$} & \colhead{$\alpha$ (median [95\% CI])}
}
\startdata
GD 362         & 10540 & \textcolor{blue}{He} & 9 & 0 & Al, Cr, Fe, Mg, Mn, Na, Ni, Si, Ti & 3.25 & 0.018 [0.008,0.032] \\
HE 0106$-$3253 & 17350 & \textcolor{blue}{H}  & 3 & 0 & Fe, Mg, Si                         & 2.01 & 0.144 [0.046,0.372] \\
GD 56          & 15270 & \textcolor{blue}{H}  & 4 & 0 & Fe, Mg, Ni, Si                     & 0.23 & 0.132 [0.022,0.382] \\
WD 1202$-$232  & 8619  & \textcolor{blue}{H}  & 1 & 0 & Fe                                 & 5.42 & 0.716 [0.348,0.984] \\
WD 2216$-$657  & 9300  & \textcolor{blue}{He} & 2 & 0 & Fe, Si                             & 2.67 & 0.100 [0.040,0.194] \\
WD 1145$+$288  & 12140 & \textcolor{blue}{H}  & 2 & 1 & Fe, Mg                             & 1.02 & 0.280 [0.042,0.748] \\
\enddata
\tablecomments{Same definitions as Table~\ref{tab:candidates_atm}, but for the \textcolor{blue}{diffusion-adjusted steady-state ratios}. \textcolor{blue}{Atm. gives the dominant atmospheric species used for the corresponding diffusion correction.} For WD 1145$+$288 the detected-panel column lists detections only; the one E10 upper limit is Si.}
\end{deluxetable*}

\FloatBarrier

\subsection{Calibration by injection--recovery}
\label{sec:results:calibration}

We calibrate the endmember-inference analysis through injection--recovery experiments that (i) draw natural compositions from the meteorite-derived reference, (ii) inject a known template fraction \(\alpha_{\rm true}\) using the same mixing model used in inference, and (iii) impose empirically matched WD detection patterns, including upper limits and heteroskedastic uncertainties.

\subsubsection{False-positive control}
\label{subsec:results:fp}

Under natural-only injections (\(\alpha_{\rm true}=0\)), the empirically measured false-positive rate for \(\mathrm{BF}>10\) is 0.011 for the photospheric track and 0.002 for the diffusion-corrected track. For the more stringent \(\mathrm{BF}>100\) threshold, the false-positive rate is 0.002 for the photospheric track and 0.0 for the diffusion-corrected track. This test is best interpreted as a statistical model-consistency check: if the data are drawn from the same natural model used in inference and are passed through the observed pattern pool, the analysis very rarely generates spurious strong evidence. The lower \texttt{acc\_ss} rate is driven mainly by its richer element coverage and slightly stricter censoring patterns, not by a different Bayes-factor construction. It is not, however, a full physical false-positive budget for time-dependent accretion or atmospheric transport outside the adopted model family.

\subsubsection{Detection power versus injected mixing fraction and dependence on element richness}
\label{subsec:results:power}

\begin{figure*}[t]
\centering
\includegraphics[width=0.85\textwidth]{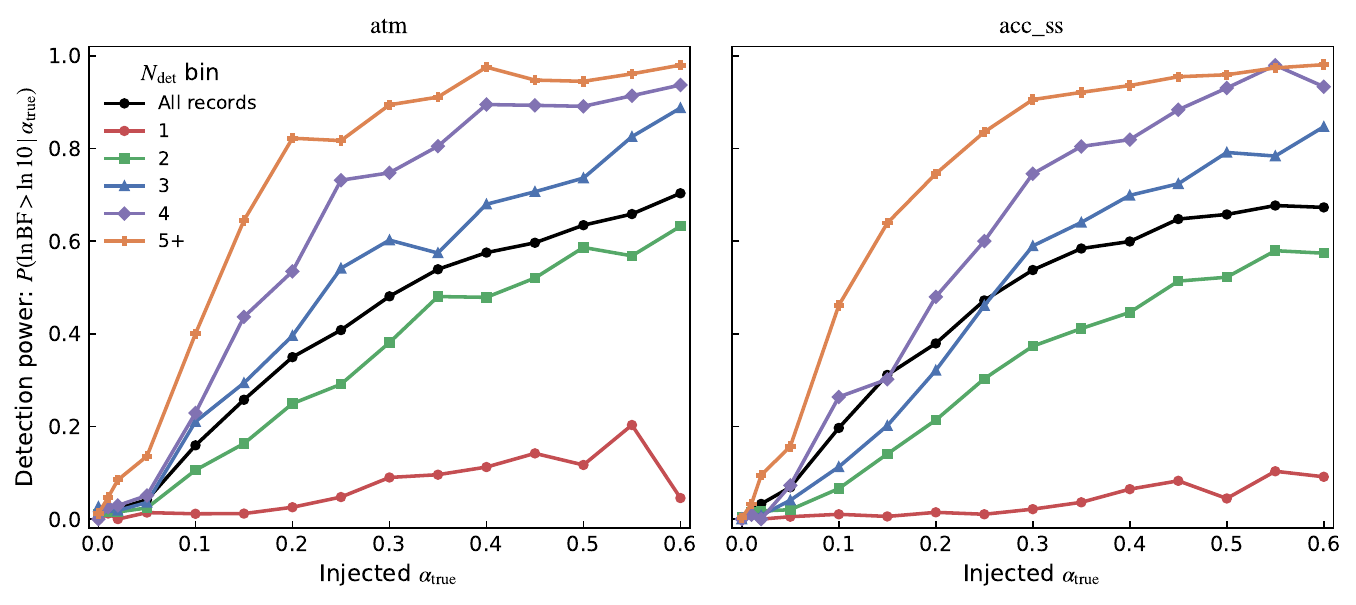}
\caption{\textbf{Injection--recovery detection power is dominated by element richness.}
Strong-evidence detection power, i.e. \(P(\ln {\rm BF}>\ln 10 \mid \alpha_{\rm true})\), is plotted against the injected Ca-normalized mixing fraction \(\alpha_{\rm true}\). Curves show recovery for all records together and after stratifying by the number of detected elements \(N_{\rm det}\). \textbf{a,} Photospheric track (\texttt{atm}). \textbf{b,} Diffusion-adjusted track (\texttt{acc\_ss}).}
\label{fig:power}
\end{figure*}

Detection power, as defined in equation~(\ref{eq:power_def}), increases with injected \(\alpha_{\rm true}\) (Fig.~\ref{fig:power}). For the \(\mathrm{BF}>10\) decision rule, the photospheric-track power is 0.159 at \(\alpha_{\rm true}=0.10\), 0.350 at \(\alpha_{\rm true}=0.20\), and 0.634 at \(\alpha_{\rm true}=0.50\). The diffusion-corrected track shows slightly higher power at moderate \(\alpha_{\rm true}\), consistent with its richer element panels: 0.197 at \(\alpha_{\rm true}=0.10\), 0.397 at \(\alpha_{\rm true}=0.20\), and 0.658 at \(\alpha_{\rm true}=0.50\).

Stratifying by \(N_{\rm det}\) shows that discrimination is driven primarily by chemical information content. At fixed \(\alpha_{\rm true}\), records with richer detected element panels achieve substantially higher power, while sparse panels rise slowly and remain well below information-rich curves at moderate \(\alpha_{\rm true}\). For records with \(N_{\rm det}\ge 5\), the \(\mathrm{BF}>10\) power at \(\alpha_{\rm true}=0.10\) is 0.400 (\texttt{atm}) and 0.461 (\texttt{acc\_ss}), rising to 0.945 and 0.959 at \(\alpha_{\rm true}=0.50\), respectively.

The exact identity of the measured elements also matters. Restricting to exact five-element panels already present in the archive, the highest-power panels repeatedly include Fe and Mg together with Cr and Ti, with Ni, Si, or Na providing the fifth discriminator (Table~\ref{tab:best_panels}). In the low-\(N_{\rm det}\) regime the literature is dominated by Fe and Mg alone, which explains why single- or two-element records can produce intriguing Bayes factors while still remaining weakly diagnostic.

\FloatBarrier

\subsubsection{Mixing-fraction recovery}
\label{subsec:results:alpha_recovery}

\begin{figure}[t]
\centering
\includegraphics[width=0.85\linewidth]{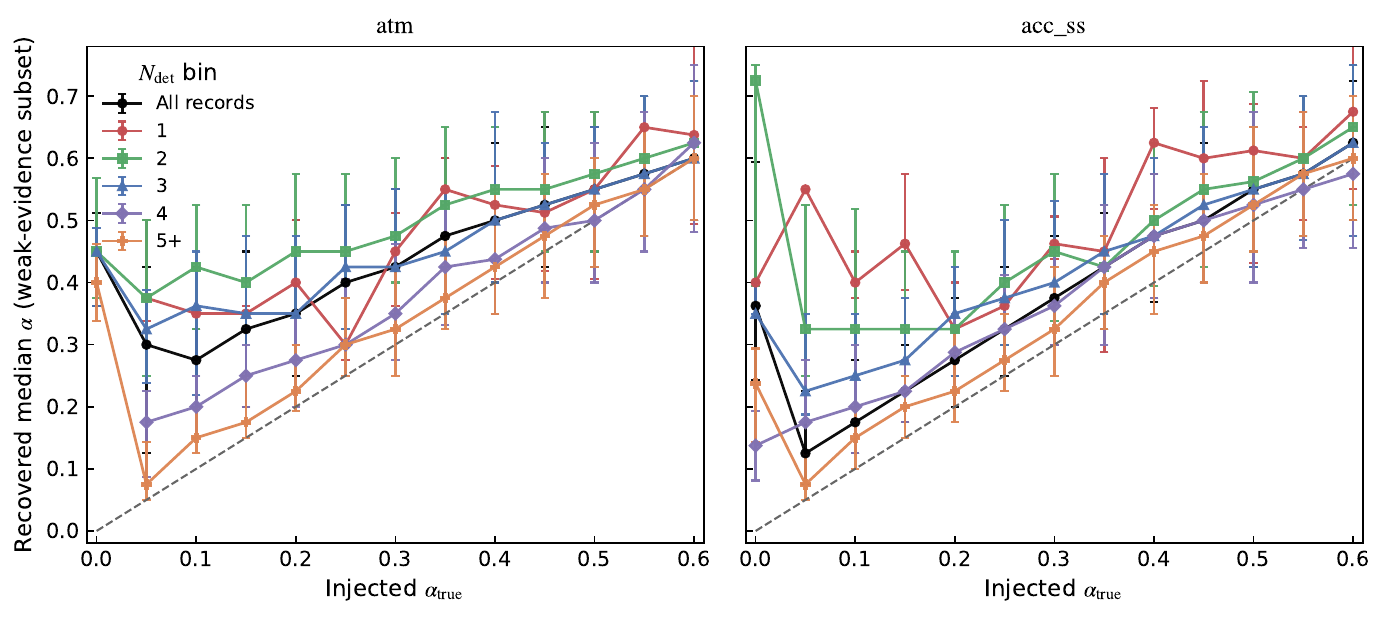}
\caption{\textbf{Calibration of mixing-fraction recovery with explicit \(N_{\rm det}\) stratification.}
Comparison of injected versus recovered Ca-normalized mixing fractions (\(\alpha\)) for synthetic records passing the weak-evidence threshold \(\ln \mathrm{BF}>1\). Curves show the median recovered slab-posterior median \(\alpha\) in bins of \(N_{\rm det}\); error bars denote the interquartile range across retained simulations. \textbf{a,} Photospheric track (\texttt{atm}). \textbf{b,} Diffusion-adjusted track (\texttt{acc\_ss}). The dashed line indicates perfect recovery.}
\label{fig:alpha_recovery}
\end{figure}

We assess recovery of the mixing fraction \(\alpha\) by comparing injected \(\alpha_{\rm true}\) to recovered posterior summaries for synthetic records passing a weak-evidence threshold (\(\ln\mathrm{BF}>1\); Fig.~\ref{fig:alpha_recovery}). The fraction of synthetic records entering this weak-evidence subset is 0.304, 0.526, and 0.777 in \texttt{atm} at \(\alpha_{\rm true}=0.10,0.20,0.50\), and 0.338, 0.522, and 0.722 in \texttt{acc\_ss}. Recovery is broadly monotonic and improves with information content, but selection on evidence introduces bias at low signal-to-noise and low \(N_{\rm det}\): records passing the threshold tend to preferentially sample upward fluctuations in apparent signal. This is why Fig.~\ref{fig:alpha_recovery} is stratified by \(N_{\rm det}\); the low-\(N_{\rm det}\) curves show the strongest positive bias at small \(\alpha_{\rm true}\), while the richer panels approach the one-to-one line much more quickly.

\FloatBarrier

\subsubsection{A natural negative control template}
\label{subsec:results:control}

\begin{figure}[t]
\centering
\includegraphics[width=0.40\linewidth]{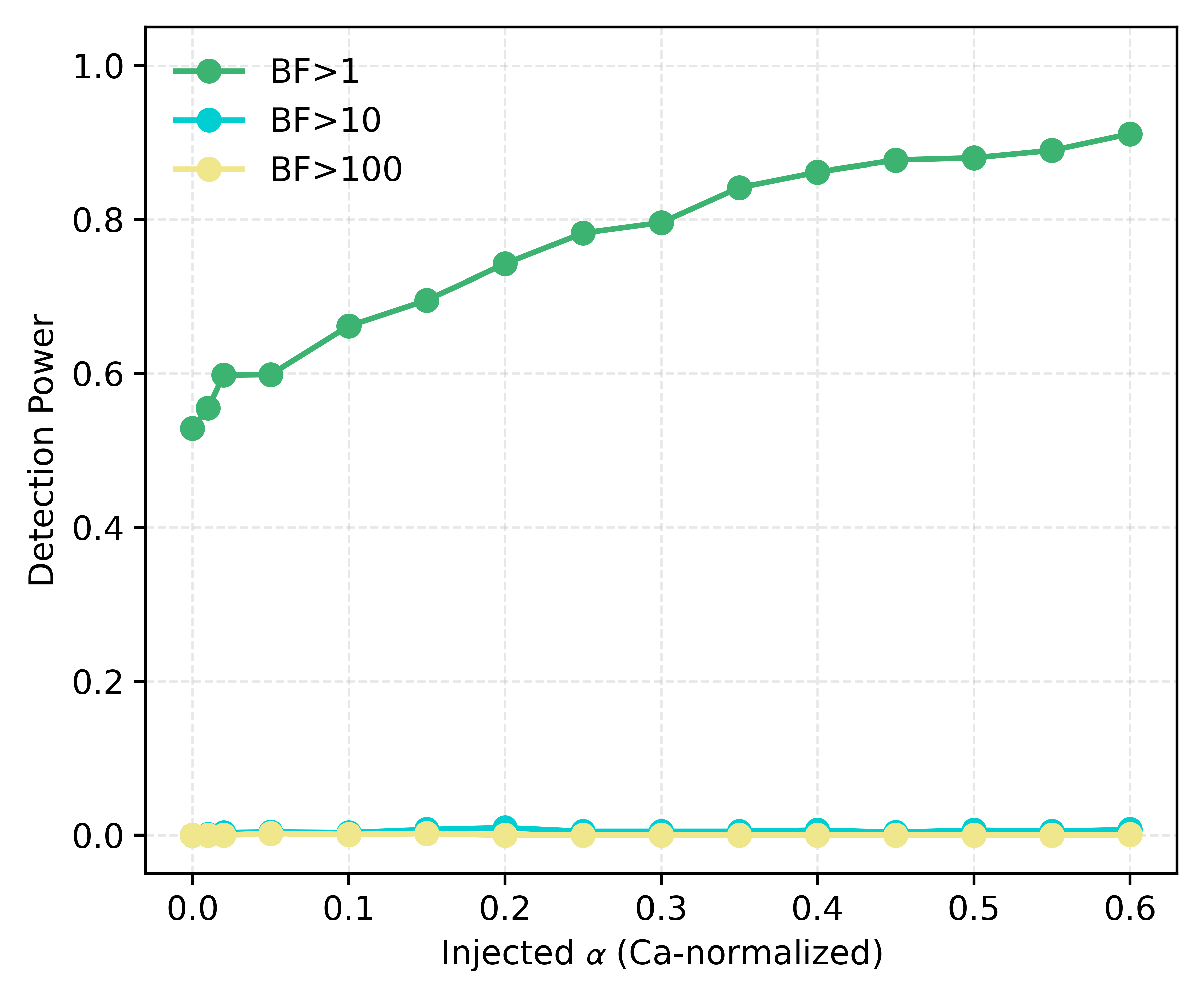}\hspace{0.3cm}
\includegraphics[width=0.40\linewidth]{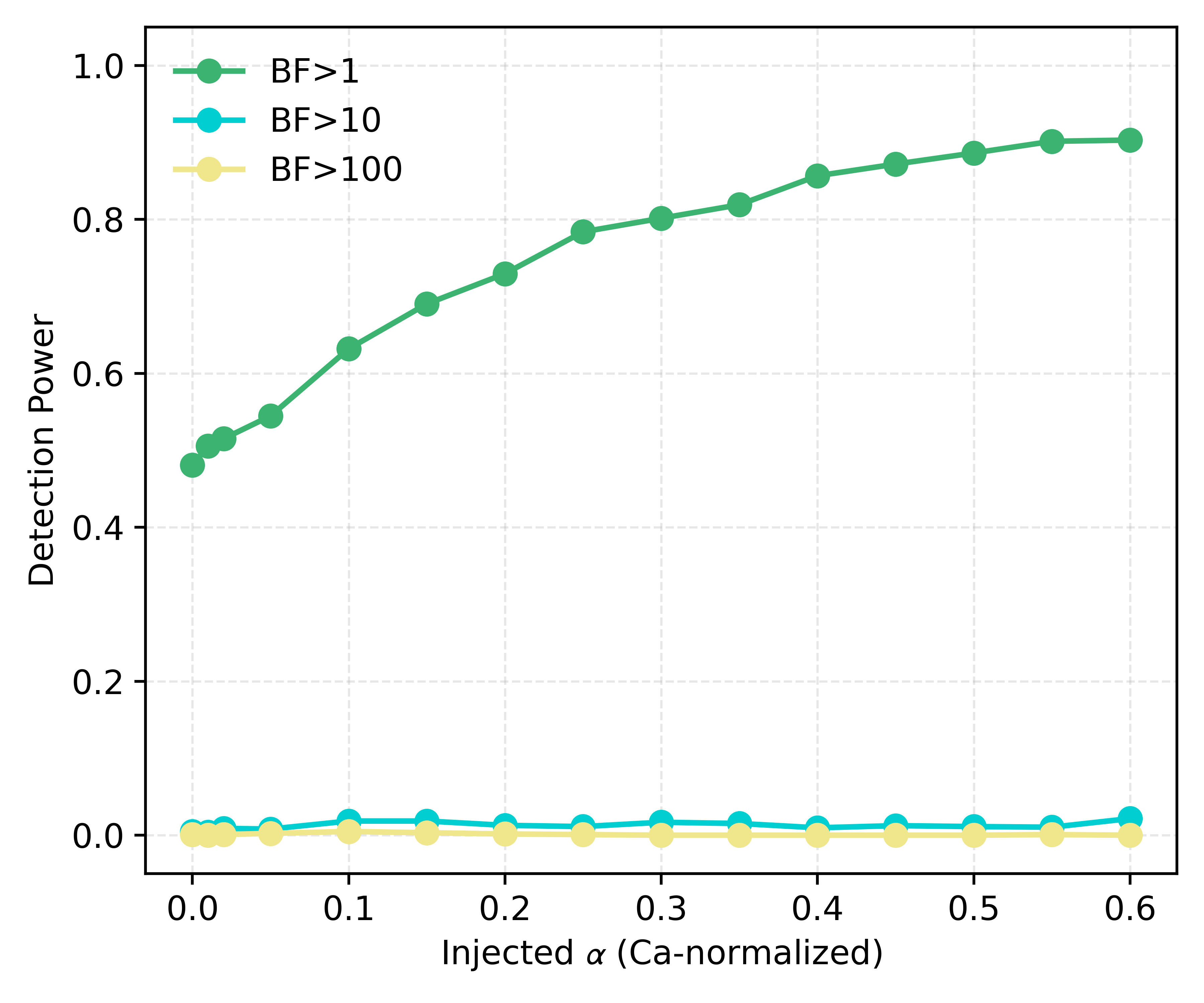}
\caption{\textbf{Detection power for a natural negative control.}
Strong-evidence detection power for the natural-control template (median chondrite composition), plotted against the injected Ca-normalized mixing fraction \(\alpha\). Because the injected template is itself natural, these curves visualize the high specificity discussed in Section~\ref{subsec:results:control}. Left: \texttt{atm}. Right: \texttt{acc\_ss}.}
\label{fig:control}
\end{figure}

Beyond the null case, we test \emph{specificity} using a natural negative-control template (median chondrite composition; Table~\ref{tab:endmembers}). Here specificity means the complement of the strong-evidence false-positive rate when the injected template is itself natural, i.e. \(1-P(\ln{\rm BF}>\ln 10)\). The same Ca-normalized linear-ratio mixing definition of equation~(\ref{eq:mix_linear}) is used here, except that \(\mathbf{R}_{\rm temp}\) is the median-chondrite vector rather than the siderophile concentrate. Across injected mixing fractions \(\alpha \in [0,0.6]\), this control yields effectively zero probability of satisfying the strong-evidence threshold (\(\mathrm{BF}>10\)) in both tracks (Fig.~\ref{fig:control}), showing that strong evidence is not a generic response to mixing with an arbitrary fixed vector.

\FloatBarrier

\subsection{Population-level detectable incidence under the fiducial processed-template hypothesis}
\label{sec:results:pi}

Using the record-level evidences, we infer the detectable-incidence parameter \(\pi\) under the prevalence model (Section~\ref{subsec:methods:prevalence}). For the photospheric track, the posterior median is \(\tilde{\pi}=0.011\) with a 95\% credible interval of \([0.003,\,0.024]\). For the diffusion-corrected track, the posterior median is \(\tilde{\pi}=0.041\) with a 95\% credible interval of \([0.005,\,0.108]\). Under the adopted template hypothesis and decision rule, these values indicate that only a small fraction of current literature records are \emph{detectably} better described by a non-zero processed component.

A separate prevalence-recovery calibration---constructed from synthetic catalogues assembled from the same record-level injections but with known input \(\pi_{\rm true}\)---shows that \(\pi\) inference can be conservative when detectability is limited by sparse element panels. The record-level posterior is also mildly \(N_{\rm det}\)-dependent: in \texttt{atm}, the \(N_{\rm det}\le 1\) bin yields \(\tilde{\pi}\approx 0.062\) with a very broad interval, whereas the \(N_{\rm det}=2\) to 5+ bins cluster between \(\approx 0.009\) and \(\approx 0.040\); the same qualitative behaviour appears in \texttt{acc\_ss} but with broader posteriors because of the smaller sample size. We therefore interpret \(\tilde{\pi}\) as a calibrated detectable-incidence constraint under current measurement patterns, rather than as a direct estimate of the true physical incidence of processed material.

\section{Discussion}
\label{sec:discussion}

\subsection{Robustness across photospheric and diffusion-corrected tracks}
\label{subsec:discussion:tracks}

We analyse both published photospheric ratios (\texttt{atm}) and \textcolor{blue}{diffusion-timescale-corrected steady-state ratios} (\texttt{acc\_ss}). The \texttt{acc\_ss} subset is smaller but generally information-richer, leading to a higher fraction of weakly positive Bayes factors. Most of the gain seen in \texttt{acc\_ss} can be traced to this change in element coverage: across matched injection--recovery curves, the typical within-\(N_{\rm det}\) difference between the two tracks is only a few $\times 10^{-2}$ in absolute power, whereas the jump between adjacent \(N_{\rm det}\) bins is much larger. Persisting candidates are therefore best understood as records that are robust to this \emph{particular} abundance-track representation, not as records proven insensitive to all diffusion or transport modelling assumptions. In practice, the strongest physical case for interpreting an \texttt{acc\_ss} candidate as near steady state would come from ongoing-accretion indicators such as a DA atmosphere, infrared excess, or circumstellar gas; our current cross-track comparison does not by itself supply that validation. They could still shift under build-up/decline accretion phases, thermohaline mixing, convective overshoot, or revised settling calculations.

\subsection{Detectable incidence versus physical incidence}
\label{subsec:discussion:pi}

The prevalence parameter \(\pi\) inferred here is explicitly a record-level \emph{detectable} incidence conditioned on the adopted template, the meteorite-trained natural reference, and the heterogeneous measurement patterns in the literature. If the true underlying population includes processed materials that do not resemble the adopted template, or if extrasolar natural compositions extend beyond Solar System meteorite diversity, then \(\pi\) will not map one-to-one onto physical incidence. Likewise, sparse panels reduce detectability and make prevalence inference conservative. Repeated records of the same star are not the dominant driver of the reported medians: when we retain only one representative record per consolidated object, the posterior medians become \(\tilde{\pi}\approx 0.017\) for \texttt{atm} and \(\approx 0.039\) for \texttt{acc\_ss}; when we combine record-level evidences within stars, they become \(\approx 0.018\) and \(\approx 0.052\), respectively. A full unique-star occurrence rate would nevertheless require a dedicated hierarchical model for repeated observations and survey selection.

\subsection{Observational requirements for stronger constraints}
\label{subsec:discussion:requirements}

Calibration by injection--recovery demonstrates that discrimination is driven primarily by chemical information content. For the fiducial siderophile-like processed class tested here, the most informative exact five-element panels repeatedly include Fe and Mg together with Cr and Ti, plus a fifth constraint from Ni, Si, or Na. The most efficient path to stronger constraints (or discovery) is therefore deeper, broader multi-element measurements for a smaller number of well-chosen targets, rather than simply increasing sample size with one--two-element records. In practical terms, surveys optimised only for Fe and Mg are valuable as screening data, but Cr/Ti coverage and at least one of Ni/Si/Na provide the clearest leverage on the fiducial technosignature hypothesis.

\subsection{Limitations and extensions}
\label{subsec:discussion:limitations}

The framework is modular. Alternative processed templates can be substituted without changing the statistical machinery, and the same end-to-end calibration can be repeated for future datasets with more uniform selection and broader element inventories. Key limitations include: (i) reliance on Solar System meteorites as the natural reference, (ii) heterogeneous systematics across literature WD analyses that motivate conservative uncertainty inflation, (iii) the fact that record-level incidence does not directly translate to unique-star occurrence without modelling repeated measurements of the same system, and (iv) the simplified treatment of atmospheric evolution in the \texttt{acc\_ss} track, which omits build-up/decline phases, thermohaline transport, and convective overshoot.

\section{Summary and Conclusions}
\label{sec:summary}

We performed a calibrated record-level search for a stylised processed chemical composition in polluted white dwarfs by comparing a meteorite-trained natural reference to a natural-plus-template mixture hypothesis across \textbf{697} archival abundance records. Strong evidence for the fiducial siderophile-enriched template is uncommon in current data: \textbf{8/697} \texttt{atm} records exceed \(\mathrm{BF}>10\) and \textbf{4/697} exceed \(\mathrm{BF}>100\), while \textbf{6/148} \texttt{acc\_ss} records exceed \(\mathrm{BF}>10\). We provide ranked candidate records and inferred mixing fractions and infer low detectable record-level incidence under a prevalence model (\(\tilde{\pi}=0.011\) for \texttt{atm} and \(\tilde{\pi}=0.041\) for \texttt{acc\_ss}). End-to-end injection--recovery calibration shows that robust discrimination is driven primarily by element richness, typically requires \(\gtrsim 5\) detected elements, and---for the fiducial siderophile test---is most strongly helped by panels containing Fe, Mg, Cr, and Ti plus Ni, Si, or Na. These calibrations define concrete observational priorities for future multi-element spectroscopy aimed at constraining (or discovering) processed composition classes in accreted debris.

\section*{Data Availability}
White-dwarf photospheric abundance measurements and associated literature provenance and metadata were obtained from the Planetary Enriched White Dwarf Database (PEWDD) \citep{Williams2024PEWDD}, available at \url{https://github.com/jamietwilliams/PEWDD}.

Whole-rock meteorite analyses used to learn the natural-composition reference were downloaded from the Astromaterials Data System Synthesis \citep{AstromatSynthesis,Hezel2025MetBaseAstromat} at \url{https://search.astromat.org/}.

White-dwarf atmospheric parameters and element diffusion timescales used for the steady-state (\texttt{acc\_ss}) transformation were obtained from the Montreal White Dwarf Database (MWDD) \citep{Dufour2017MWDD} at \url{https://www.montrealwhitedwarfdatabase.org/}, using the “Evolutionary models and diffusion timescales” interface \url{https://www.montrealwhitedwarfdatabase.org/evolution.html}.

\begin{acknowledgments}
This work was supported by the National Key R\&D Program of China (No.\ 2024YFA1611804),
the China Manned Space Program (CMS-CSST2025-A01), the National SKA Program of China under Grant No. 2025SKA0120104, the Shandong Provincial Natural Science Foundation (ZR2024QA180),
and the Scientific Research Fund of Dezhou University (4022504019). We thank the maintainers and contributors of PEWDD for compiling and standardising polluted white dwarf abundance measurements, the Astromaterials Data System (Astromat) team for providing accessible laboratory geochemistry holdings, and the Montreal White Dwarf Database team for diffusion-timescale and model resources.
\end{acknowledgments}

\clearpage
\appendix
\restartappendixnumbering
\FloatBarrier

\begin{figure}[!htbp]
\centering
\includegraphics[width=0.8\linewidth]{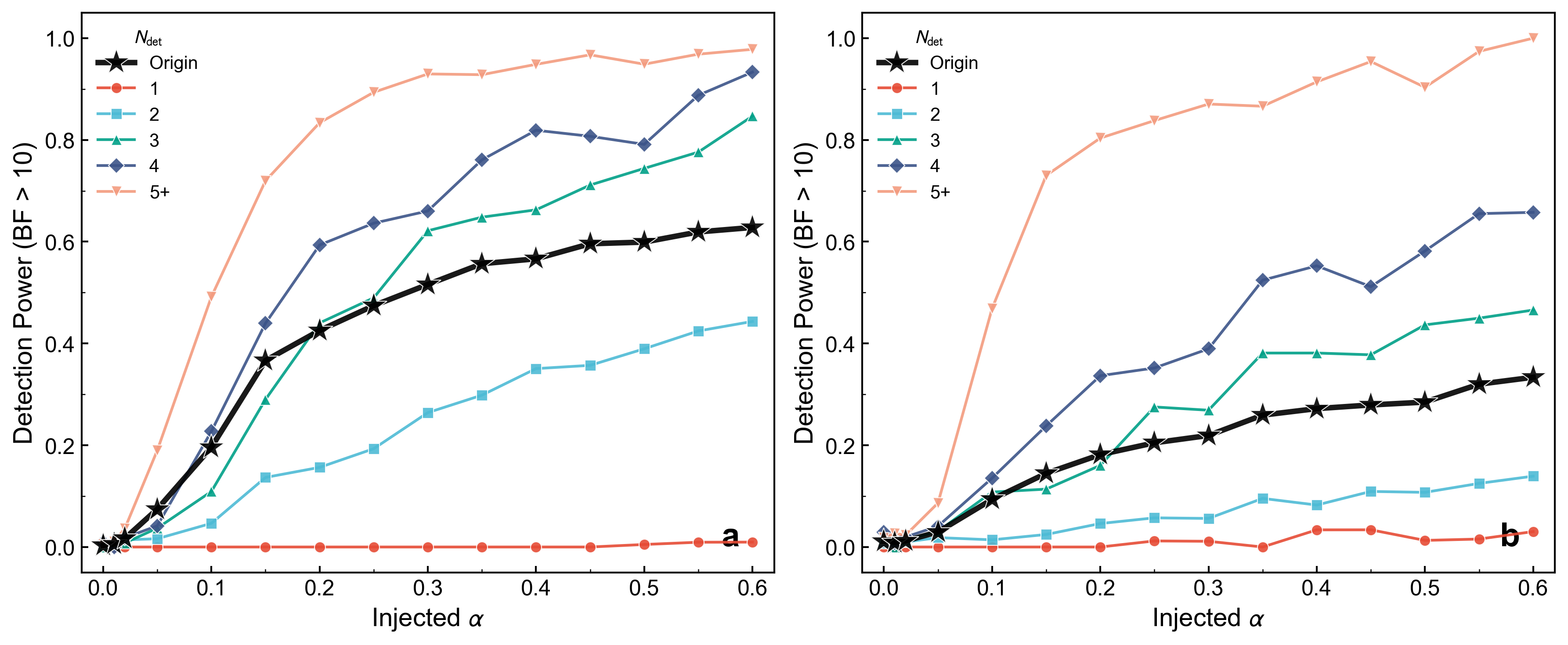}
\caption{\textbf{Detection power for the refractory-silicate concentrate template.}
Probability of recovering strong evidence (\(\mathrm{BF}>10\)) versus injected \(\alpha_{\rm true}\), stratified by the number of detected elements \(N_{\mathrm{det}}\).
\textbf{a,} \texttt{acc\_ss}. \textbf{b,} \texttt{atm}.}
\label{fig:supp_power_refractory}
\end{figure}

\begin{figure}[!htbp]
\centering
\includegraphics[width=0.8\linewidth]{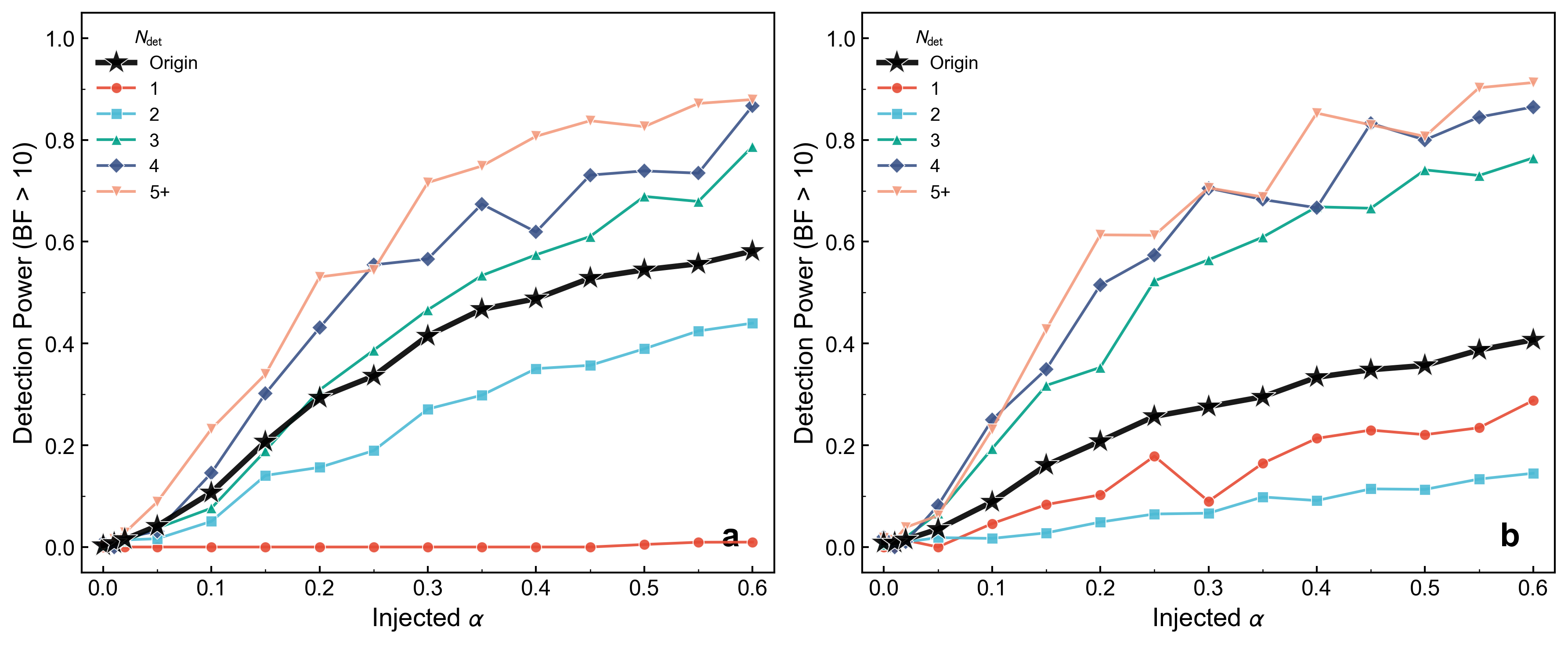}
\caption{\textbf{Detection power for the alkali-aluminosilicate concentrate template.}
Same format as Fig.~\ref{fig:supp_power_refractory}. \textbf{a,} \texttt{acc\_ss}. \textbf{b,} \texttt{atm}.}
\label{fig:supp_power_alkali}
\end{figure}

\begin{figure}[!htbp]
\centering
\includegraphics[width=0.8\linewidth]{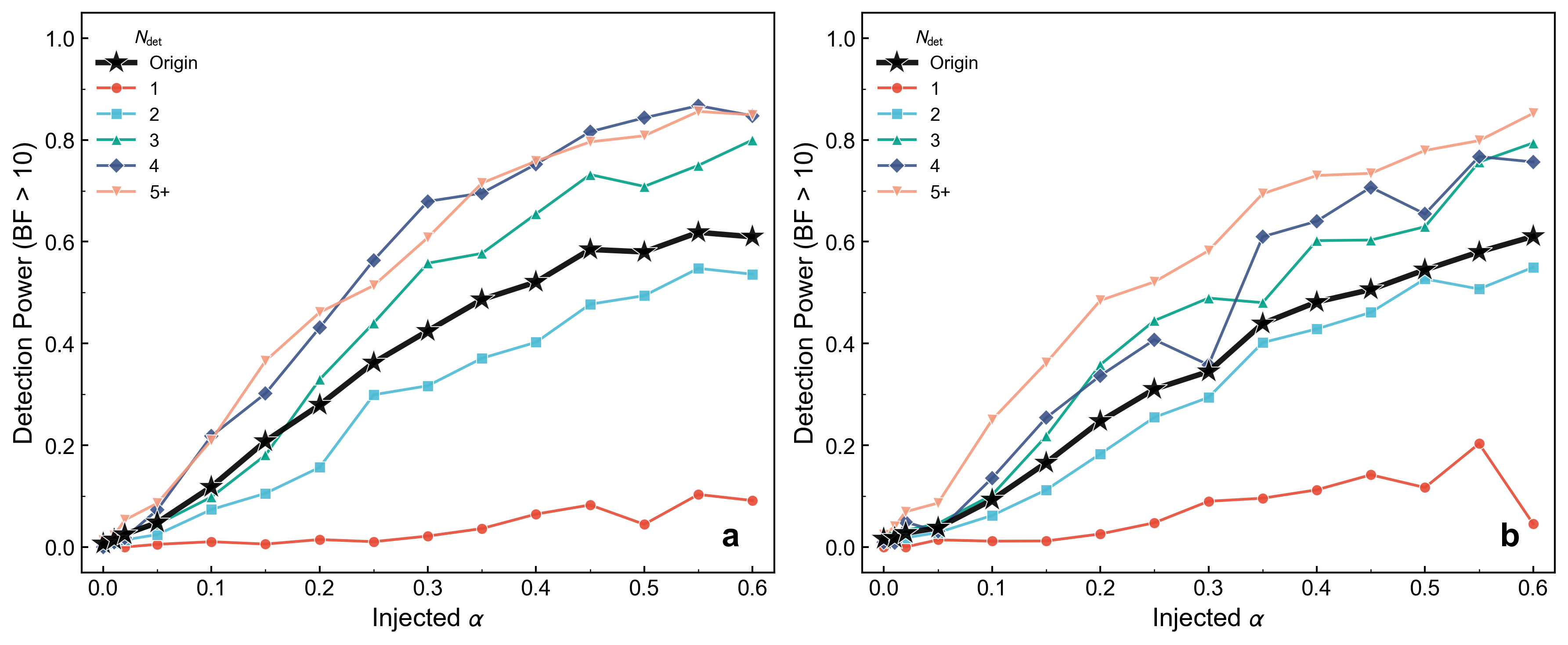}
\caption{\textbf{Detection power for the sulfide-metal concentrate template.}
Same format as Fig.~\ref{fig:supp_power_refractory}. \textbf{a,} \texttt{acc\_ss}. \textbf{b,} \texttt{atm}.}
\label{fig:supp_power_sulfide}
\end{figure}

\begin{figure}[!htbp]
\centering
\includegraphics[width=0.8\linewidth]{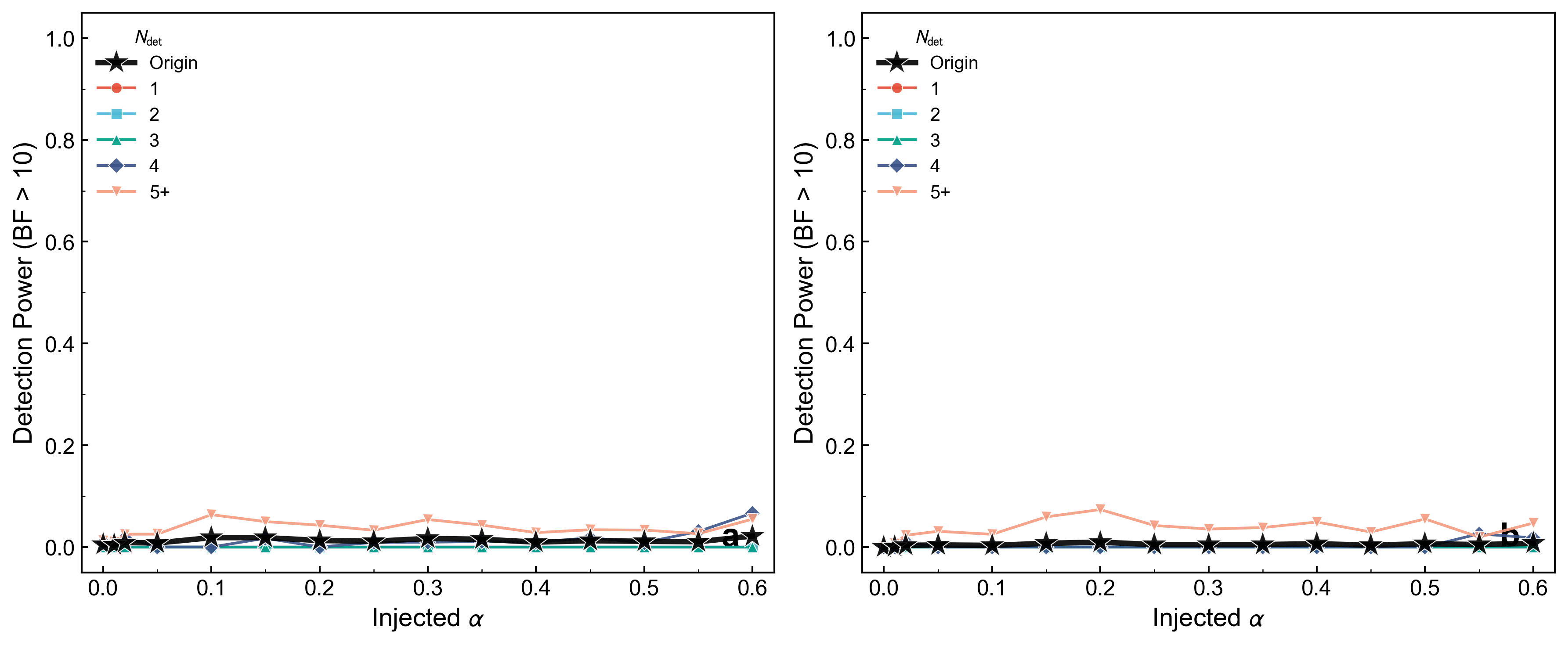}
\caption{\textbf{Detection power for a natural negative-control template (median chondrite) with \(N_{\mathrm{det}}\) stratification.}
Same format as Fig.~\ref{fig:supp_power_refractory}. Strong-evidence detection power remains near zero across injected \(\alpha_{\rm true}\) and \(N_{\mathrm{det}}\).
\textbf{a,} \texttt{acc\_ss}. \textbf{b,} \texttt{atm}.}
\label{fig:supp_power_natural_control}
\end{figure}

Appendix~A summarises the fixed template (endmember) compositions used throughout this work in Ca-normalized
log mass-ratio space, including the fiducial siderophile-enriched template, the natural negative control (median
chondrite), and the additional stylised processed templates used for sensitivity tests (Table~\ref{tab:endmembers}).
Appendix~B presents supplementary end-to-end injection--recovery detection-power curves for the alternative
processed templates---refractory-silicate concentrate (Fig.~\ref{fig:supp_power_refractory}), alkali-aluminosilicate concentrate (Fig.~\ref{fig:supp_power_alkali}),
and sulfide-metal concentrate (Fig.~\ref{fig:supp_power_sulfide})---and for the $N_{\mathrm{det}}$-stratified natural negative control
(median chondrite; Fig.~\ref{fig:supp_power_natural_control}). These appendix materials therefore both document the template parameter choices
and show how detectability depends on the assumed processed-composition class beyond the main-text calibration.
\par\medskip

\section{Template definitions used in this work}
\label{app:endmembers}

\begin{deluxetable}{lrrrrr}
\tablecaption{\textbf{Template (endmember) definitions used in this work.}\label{tab:endmembers}}
\tabletypesize{\scriptsize}
\tablewidth{0pt}
\tablehead{
\colhead{Element ($Z$)} &
\colhead{Natural Control} &
\colhead{Siderophile Conc.} &
\colhead{Refractory Silicate} &
\colhead{Alkali Aluminosil.} &
\colhead{Sulfide Metal}
}
\startdata
Al & $-0.017$ & $-0.364$ & $1.392$  & $0.992$  & $-0.033$ \\
Cr & $-0.553$ & $0.788$  & $-1.888$ & $-1.888$ & $-0.622$ \\
Fe & $1.248$  & $2.359$  & $-0.321$ & $-0.321$ & $2.359$ \\
Mg & $1.062$  & $-0.401$ & $1.027$  & $1.027$  & $1.027$ \\
Mn & $-0.727$ & $0.473$  & $-2.043$ & $-0.787$ & $-0.787$ \\
Na & $-0.295$ & $-1.675$ & $-0.403$ & $1.040$  & $-0.403$ \\
Ni & $0.013$  & $1.172$  & $-0.886$ & $-0.886$ & $1.172$ \\
P  & $-0.960$ & $-1.279$ & $-1.072$ & $-1.072$ & $-1.072$ \\
S  & $-0.060$ & $-0.148$ & $0.560$  & $0.560$  & $1.904$ \\
Si & $1.161$  & $0.359$  & $2.278$  & $2.278$  & $1.123$ \\
Ti & $-1.284$ & $-1.808$ & $0.289$  & $-1.294$ & $-1.294$ \\
\enddata
\tablecomments{\textbf{Endmember (template) definitions used in this work.}
The \textbf{Natural Control} represents the median composition of chondrites and is used as a negative control. The \textbf{Siderophile Concentrate} is the fiducial processed endmember used for record-level inference in the main text and should be interpreted as a stylised metal-rich, silicate-poor component rather than as a unique industrial product. The \textbf{Refractory Silicate}, \textbf{Alkali Aluminosilicate}, and \textbf{Sulfide Metal} endmembers are additional stylised processed hypotheses used to assess detectability in injection--recovery tests. Processed endmembers are constructed \emph{a priori} by setting enriched elements to the 99\textsuperscript{th} quantile of the meteorite distribution plus 0.6--1.0 dex, depleted elements to the 5\textsuperscript{th} quantile, and neutral elements to the median (50\textsuperscript{th} quantile). All values are Ca-normalized log mass ratios, \(r_Z=\log_{10}(m_Z/m_{\mathrm{Ca}})\).}
\end{deluxetable}

\FloatBarrier

\section{Supplementary injection--recovery power curves for alternative processed templates}
\label{app:supp}
\FloatBarrier

\FloatBarrier
\clearpage

\bibliography{sample701}{}
\bibliographystyle{aasjournalv7}
\end{CJK*}
\end{document}